\documentstyle[12pt,epsfig]{article}
\begin{document}
\renewcommand{\baselinestretch}{1.1}
\small\normalsize
\renewcommand{\theequation}{\arabic{section}.\arabic{equation}}
\renewcommand{\thesection}{\arabic{section}.}
\language0


\begin{flushright}
{\sc BUHEP}-95-01\\January 1995
\end{flushright}
\vspace{.1cm}

\vspace*{1.5cm}

\begin{center}

{\Large \bf Monopole currents and Dirac sheets\\in U(1) lattice gauge theory}

\vspace*{0.8cm}

{\bf Werner Kerler$^a$, Claudio Rebbi$^b$, Andreas Weber$^a$}

\vspace*{0.3cm}

{\sl $^a$ Fachbereich Physik, Universit\"at Marburg, D-35032 Marburg, 
Germany\\ 
$^b$ Department of Physics, Boston University, Boston, MA 02215, USA}
\hspace*{3.6mm}

\end{center}

\vspace*{1.5cm}

\begin{abstract} 
We show that the phases of the 4-dimensional compact U(1) lattice gauge 
theory are unambiguously characterized by the topological properties of
minimal Dirac sheets as well as of monopole currents lines. We obtain
the minimal sheets by a simulated-annealing procedure. Our results
indicate that the equivalence classes of sheet structures are the
physical relevant quantities and that intersections are not
important. In conclusion we get a percolation-type view of the phases
which holds beyond the particular boundary conditions used.

\end{abstract}

\newpage

\section{Introduction}
\hspace{3mm}
We investigate the 4-dimensional compact U(1) lattice gauge theory 
with Wilson action supplemented by a monopole term \cite{bss85}:
\begin{displaymath}
S=\beta \sum_{\mu>\nu,x} (1-\cos \Theta_{\mu\nu,x})+
\lambda \sum_{\rho,x} |M_{\rho,x}| \; , \nonumber
\end{displaymath}
where $M_{\rho,x} = \epsilon_{\rho\sigma\mu\nu}
(\bar{\Theta}_{\mu\nu,x+\sigma}-\bar{\Theta}_{\mu\nu,x})/4\pi$ and 
the physical flux $\bar{\Theta}_{\mu\nu,x}\in [-\pi,\pi)$  
is given by 
$\Theta_{\mu\nu,x}=\bar{\Theta}_{\mu\nu,x}+2\pi n_{\mu\nu,x}$
\cite{dt80}.
We consider periodic boundary conditions except for the discussion 
in Section 5 where we extend our study to a system with open boundary 
conditions.

As is well known, the system has two phases, separated for $\lambda=0$
by a first order phase transition.  The strength of this transition
decreases with increasing $\lambda$ until the transition ultimately
becomes of second order \cite{krw94}.  We have used this property to
set up a very efficient algorithm \cite{krw94,krw94a} for Monte Carlo
simulations in which $\lambda$ becomes a dynamical variable. The
$\lambda$ dependence has also allowed us to study the dynamics of the
transition in detail \cite{krw94}.

We argue that the analysis of the topological structure of the
configurations is a valuable tool for studying the phase structure,
a tool which can provide much more detailed information than just the
measurement of global observables. In the present work we use it by
analyzing monopole currents and Dirac sheets.

For individual loops the topological characterization is
straightforward, but this becomes less trivial when loops are
entangled in networks of monopole currents. Recently we have been able
to produce a mathematically sound characterization of the latter
\cite{krw94}.  This is not only necessary for the unambiguous
identification of their physical features but also for the analysis of
huge networks by computer.

Here the analysis of configurations is extended to Dirac sheets and the fact
that appropriately specified topological structures signal the phases is 
confirmed in more detail. In conclusion we point out a general principle
for characterizing the phases, which does not rely on a particular choice 
of boundary conditions.

\section{Dual-lattice structures}
\hspace{3mm}
On the dual lattice the current $J_{\rho,x}=M_{\rho,x+\rho}$, defined over the
links, obeys the conservation law $\sum_{\rho}(J_{\rho,x}-J_{\rho,x-\rho})=0$.
Current lines are defined in terms of the current as follows:
for $J_{\rho,x}=0$ there is no line on the link, 
for $J_{\rho,x}=\pm 1$ there is one line, and for $J_{\rho,x}=\pm 2$ there are
two lines, in positive or negative direction, respectively. Networks of 
currents are connected sets of current lines. For a network {\bf N} 
disconnected from the rest one can define a net current flow $\vec{f}$, with 
components $ f_{\mu_3} = \sum_{x_{\mu_0}x_{\mu_1}x_{\mu_2}}J_{\mu_3,x}$,
$J_{\mu,x}\in \mbox{{\bf N}}$. 

The Dirac string content of the plaquettes on the dual lattice is
described in terms of a variable
$p_{\rho\sigma,x}=-\frac{1}{2}\epsilon_{\rho\sigma\mu\nu}n_{\mu\nu,x+\rho+
\sigma}$, which satisfies the field equation
$\sum_{\sigma}(p_{\rho\sigma,x}-p_{\rho\sigma,x-\sigma})= J_{\rho,x}$.
To every plaquette in the dual lattice we will associate no Dirac
plaquette if $p_{\rho\sigma,x} = 0$, one Dirac plaquette if
$p_{\rho\sigma,x} = \pm 1$, with similar or opposite orientation
according to the sign of $p$, and two Dirac plaquettes if
$p_{\rho\sigma,x} = \pm 2$.

Dirac sheets are formed by connecting Dirac plaquettes with a common edge and 
appropriate orientation, so that $J_{\rho,x}=0$ except at the boundaries of the
sheets. Dirac sheet structures are not gauge invariant. However, they belong 
to equivalence classes which cannot be deformed into each other by gauge 
transformations. For given boundaries there are topologically distinct 
possibilities for the sheet structures. This is illustrated in Fig.~1 by a
simple example in 2 dimensions, where a topologically nontrivial current 
network may be accompanied by a trivial or a nontrivial Dirac sheet. The 
equivalence classes of Dirac sheets thus carry more information than 
the related current networks. 

We represent the equivalence classes of the sheet structures by their members 
with minimal area, which are obtained by minimizing the number of Dirac 
plaquettes by a gauge transformation. This permits a unique specification and,
in view of the large initial number of Dirac plaquettes, is necessary for
a computational analysis.

To address the connectedness of sheet structures, we construct minimal Dirac 
sheets by making first the connections where only two Dirac plaquettes meet at 
an edge with appropriate orientation. Then we consider the places where more 
than two Dirac plaquettes meet at an edge. If the sheets can be 
deformed (by a gauge transformation) in such a way that no more than
two plaqettes meet we connect the respective pairs. Otherwise all must 
to be connected.

\begin{figure}[tb]
\begin{center}
\leavevmode
\psfig{figure=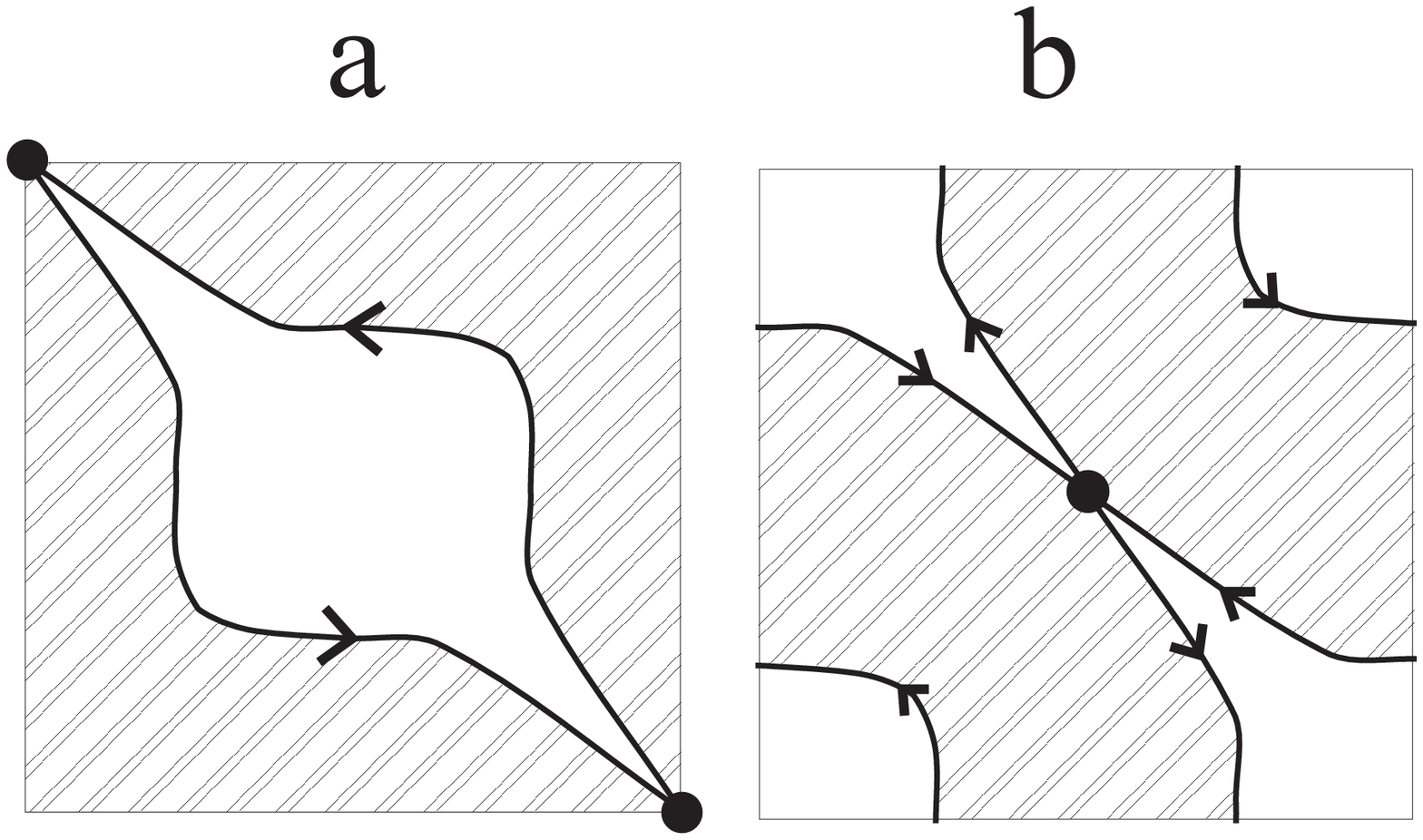,bbllx=2089bp,bblly=2284bp,bburx=2744bp,%
       bbury=2834bp,width=12cm,angle=0}
\end{center}
\caption{Example of current network with 
two possibilities for a Dirac sheet
on a 2-dimensional lattice with periodic boundary conditions.
The shaded (unshaded) area represents a topologically non-trivial
(trivial) sheet with given boundary.
(b) shifted against (a) by $L/2$ in both directions.}
\end{figure}

In Fig.~2 we use some simple 3-dimensional examples to illustrate
typical possibilities for the intersection of Dirac sheets or for the
meeting of more than two plaquettes.  In Fig.~2a, deforming one of the
sheets by a gauge transformation, the sheets can be
separated. Consequently there is no connection. The same holds for
Fig.~2b, where one must realize that common sites are not considered
as a connection of sheets. In Fig.~2c, although one can identify two
sheets, the structure cannot be separated by a gauge transformation
and, therefore, is to be considered as connected. In Fig.~2d
separation is obviously not possible.

\begin{figure}[tb]
\begin{center}
\leavevmode
\psfig{figure=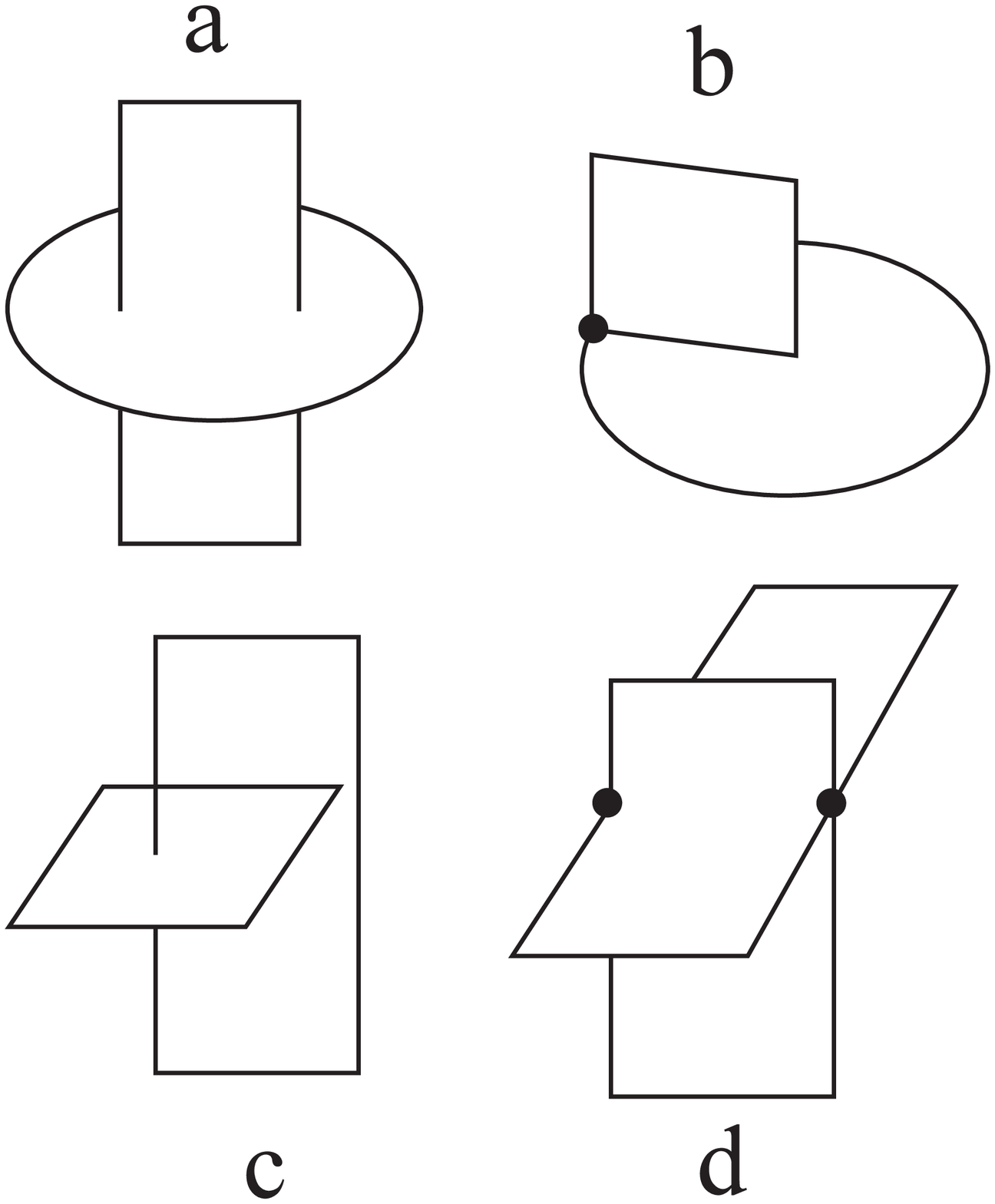,bbllx=2107bp,bblly=2173bp,bburx=2577bp,%
       bbury=2782bp,width=10cm,angle=0}
\end{center}
\caption{Typical possibilites for intersections of Dirac sheets.}
\end{figure}
\clearpage

\section{Topological analysis}
\hspace{3mm}
The elements of the fundamental homotopy group $\pi_1(\mbox{{\bf
X}},b)$ of a space {\bf X} with base point $b$ are equivalence classes
of paths starting and ending at $b$ which can be deformed continuously
into each other. Its generators may be obtained embedding a
sufficiently dense network {\bf N} into {\bf X} and performing
suitable transformations which preserve homotopy. We observe that, if
a given network {\bf N} does not wrap around in all directions, then only the
generators of a subgroup are produced.  This provides an unambiguous
characterization of networks.

Choosing one vertex point of {\bf N} to be the base point $b$ and
considering all paths which start and end at $b$, we note that a
mapping which shrinks one edge to zero length preserves the homotopy
of all of these paths. By a sequence of such mappings we can then
shift all other vertices to $b$ without changing the group content
until we finally obtain a bouquet of paths which all start and end at
$b$.

Describing a path by a vector which is the sum of oriented steps along
the path, for {\bf N} with $K_0$ vertices and $K_1$ edges the bouquet
has $K=K_1 - K_0 +1$ loops, represented by vectors $\vec{s}_i$, with
$i=1,\ldots,K$ and j-th components $s_{ij}=w_{ij} L_j$, where $L_j$ is
the lattice size. The bouquet matrix $w_{ij}$ must then be analyzed
for the content of generators of $\pi_1(\mbox{{\bf T}}^4,b)=\mbox{{\bf
Z}}^4$.

Current networks have the additional properties of path orientation
and current conservation. The maps reducing the bouquet matrix
$w_{ij}$, therefore, in addition to homotopy have to respect current
conservation. This leads to a modified Gauss elimination procedure,
whereby the addition of a row to another requires its simultaneous
subtraction from a further row. In this way one arrives at the minimal
form, with rows $\vec{a}_1$, $\ldots$, $\vec{a}_r$, $\vec{t}$,
$\vec{0}$,$\ldots$, $\vec{0}$ where $r\le 4$. The significance of this
minimal matrix becomes obvious if one switches to the pair form with
rows $\vec{a}_1$, $-\vec{a}_1$, $\ldots$, $\vec{a}_r$, $-\vec{a}_r$,
$\vec{f}$, $\vec{0}$, $\ldots$, $\vec{0}$, which exhibits the relation
to the net current flow $\vec{f}$ explicitely. The number of
independent pairs determines the number of nontrivial directions.

For the topological analysis of minimal Dirac sheets we use the
network of plaquettes as a sufficiently dense auxiliary network. The
bouquet matrix then is obtained as described before. The reduction of
the bouquet matrix is simpler here because the usual Gauss elimination
applies, which gives the minimal form with rows $\vec{a}_1$, $\ldots$,
$\vec{a}_r$, $\vec{0}$,$\ldots$, $\vec{0}$. Now the number of
independent vectors corresponds to the number of nontrivial
directions.

\clearpage
\section{Numerical results}
\hspace{3mm}
Fig.~3 shows our results for the probability $P_{\mbox{\scriptsize
net}}$ to find a current network which is nontrivial in four
directions as function of $\lambda$, with $\beta$ chosen at the
corresponding phase transition points \cite{krw94,krw94a}. For larger
$L$ or negative $\lambda$, where the peaks of the energy distribution
related to the phases are well separated \cite{krw94,krw94a},
$P_{\mbox{\scriptsize net}}$ is seen to be very close to 1 for the hot
(confining) phase, and very close to 0 for the cold (Coulomb) phase.
We thus see that the two phases have an unambiguous topological
characterization, provided by the existence of a nontrivial network in
the hot phase and its absence in the cold phase.

\begin{figure}
\setlength{\unitlength}{0.240900pt}
\ifx\plotpoint\undefined\newsavebox{\plotpoint}\fi
\sbox{\plotpoint}{\rule[-0.200pt]{0.400pt}{0.400pt}}%
\begin{picture}(1349,900)(0,0)
\font\gnuplot=cmr10 at 10pt
\gnuplot
\sbox{\plotpoint}{\rule[-0.200pt]{0.400pt}{0.400pt}}%
\put(220.0,113.0){\rule[-0.200pt]{4.818pt}{0.400pt}}
\put(198,113){\makebox(0,0)[r]{0.0}}
\put(1265.0,113.0){\rule[-0.200pt]{4.818pt}{0.400pt}}
\put(220.0,244.0){\rule[-0.200pt]{4.818pt}{0.400pt}}
\put(198,244){\makebox(0,0)[r]{0.2}}
\put(1265.0,244.0){\rule[-0.200pt]{4.818pt}{0.400pt}}
\put(220.0,374.0){\rule[-0.200pt]{4.818pt}{0.400pt}}
\put(198,374){\makebox(0,0)[r]{0.4}}
\put(1265.0,374.0){\rule[-0.200pt]{4.818pt}{0.400pt}}
\put(220.0,505.0){\rule[-0.200pt]{4.818pt}{0.400pt}}
\put(198,505){\makebox(0,0)[r]{0.6}}
\put(1265.0,505.0){\rule[-0.200pt]{4.818pt}{0.400pt}}
\put(220.0,636.0){\rule[-0.200pt]{4.818pt}{0.400pt}}
\put(198,636){\makebox(0,0)[r]{0.8}}
\put(1265.0,636.0){\rule[-0.200pt]{4.818pt}{0.400pt}}
\put(220.0,767.0){\rule[-0.200pt]{4.818pt}{0.400pt}}
\put(198,767){\makebox(0,0)[r]{1.0}}
\put(1265.0,767.0){\rule[-0.200pt]{4.818pt}{0.400pt}}
\put(332.0,113.0){\rule[-0.200pt]{0.400pt}{4.818pt}}
\put(332,68){\makebox(0,0){-0.3}}
\put(332.0,812.0){\rule[-0.200pt]{0.400pt}{4.818pt}}
\put(668.0,113.0){\rule[-0.200pt]{0.400pt}{4.818pt}}
\put(668,68){\makebox(0,0){0.0}}
\put(668.0,812.0){\rule[-0.200pt]{0.400pt}{4.818pt}}
\put(1005.0,113.0){\rule[-0.200pt]{0.400pt}{4.818pt}}
\put(1005,68){\makebox(0,0){0.3}}
\put(1005.0,812.0){\rule[-0.200pt]{0.400pt}{4.818pt}}
\put(220.0,113.0){\rule[-0.200pt]{256.559pt}{0.400pt}}
\put(1285.0,113.0){\rule[-0.200pt]{0.400pt}{173.207pt}}
\put(220.0,832.0){\rule[-0.200pt]{256.559pt}{0.400pt}}
\put(45,472){\makebox(0,0){$P_{net}$ }}
\put(752,23){\makebox(0,0){$\lambda$}}
\put(752,877){\makebox(0,0){ }}
\put(220.0,113.0){\rule[-0.200pt]{0.400pt}{173.207pt}}
\put(668,767){\makebox(0,0){$\times$}}
\put(781,767){\makebox(0,0){$\times$}}
\put(893,701){\makebox(0,0){$\times$}}
\put(1005,675){\makebox(0,0){$\times$}}
\put(668,113){\makebox(0,0){$\times$}}
\put(781,113){\makebox(0,0){$\times$}}
\put(893,113){\makebox(0,0){$\times$}}
\put(1005,126){\makebox(0,0){$\times$}}
\put(668.0,760.0){\rule[-0.200pt]{0.400pt}{3.132pt}}
\put(658.0,760.0){\rule[-0.200pt]{4.818pt}{0.400pt}}
\put(658.0,773.0){\rule[-0.200pt]{4.818pt}{0.400pt}}
\put(781.0,760.0){\rule[-0.200pt]{0.400pt}{3.132pt}}
\put(771.0,760.0){\rule[-0.200pt]{4.818pt}{0.400pt}}
\put(771.0,773.0){\rule[-0.200pt]{4.818pt}{0.400pt}}
\put(893.0,669.0){\rule[-0.200pt]{0.400pt}{15.658pt}}
\put(883.0,669.0){\rule[-0.200pt]{4.818pt}{0.400pt}}
\put(883.0,734.0){\rule[-0.200pt]{4.818pt}{0.400pt}}
\put(1005.0,642.0){\rule[-0.200pt]{0.400pt}{15.899pt}}
\put(995.0,642.0){\rule[-0.200pt]{4.818pt}{0.400pt}}
\put(995.0,708.0){\rule[-0.200pt]{4.818pt}{0.400pt}}
\put(668.0,113.0){\rule[-0.200pt]{0.400pt}{1.686pt}}
\put(658.0,113.0){\rule[-0.200pt]{4.818pt}{0.400pt}}
\put(658.0,120.0){\rule[-0.200pt]{4.818pt}{0.400pt}}
\put(781.0,113.0){\rule[-0.200pt]{0.400pt}{3.132pt}}
\put(771.0,113.0){\rule[-0.200pt]{4.818pt}{0.400pt}}
\put(771.0,126.0){\rule[-0.200pt]{4.818pt}{0.400pt}}
\put(893.0,113.0){\rule[-0.200pt]{0.400pt}{4.818pt}}
\put(883.0,113.0){\rule[-0.200pt]{4.818pt}{0.400pt}}
\put(883.0,133.0){\rule[-0.200pt]{4.818pt}{0.400pt}}
\put(1005.0,113.0){\rule[-0.200pt]{0.400pt}{7.950pt}}
\put(995.0,113.0){\rule[-0.200pt]{4.818pt}{0.400pt}}
\put(995.0,146.0){\rule[-0.200pt]{4.818pt}{0.400pt}}
\put(332,124){\circle{12}}
\put(388,125){\circle{12}}
\put(444,116){\circle{12}}
\put(500,120){\circle{12}}
\put(556,130){\circle{12}}
\put(612,117){\circle{12}}
\put(668,120){\circle{12}}
\put(724,122){\circle{12}}
\put(781,125){\circle{12}}
\put(837,121){\circle{12}}
\put(893,117){\circle{12}}
\put(949,119){\circle{12}}
\put(1005,126){\circle{12}}
\put(1061,118){\circle{12}}
\put(1117,114){\circle{12}}
\put(1173,122){\circle{12}}
\put(332,759){\circle{12}}
\put(388,753){\circle{12}}
\put(444,743){\circle{12}}
\put(500,726){\circle{12}}
\put(556,713){\circle{12}}
\put(612,680){\circle{12}}
\put(668,658){\circle{12}}
\put(724,639){\circle{12}}
\put(781,619){\circle{12}}
\put(837,606){\circle{12}}
\put(893,565){\circle{12}}
\put(949,537){\circle{12}}
\put(1005,527){\circle{12}}
\put(1061,478){\circle{12}}
\put(1117,399){\circle{12}}
\put(1173,383){\circle{12}}
\put(332.0,118.0){\rule[-0.200pt]{0.400pt}{2.891pt}}
\put(322.0,118.0){\rule[-0.200pt]{4.818pt}{0.400pt}}
\put(322.0,130.0){\rule[-0.200pt]{4.818pt}{0.400pt}}
\put(388.0,117.0){\rule[-0.200pt]{0.400pt}{3.854pt}}
\put(378.0,117.0){\rule[-0.200pt]{4.818pt}{0.400pt}}
\put(378.0,133.0){\rule[-0.200pt]{4.818pt}{0.400pt}}
\put(444.0,113.0){\rule[-0.200pt]{0.400pt}{1.445pt}}
\put(434.0,113.0){\rule[-0.200pt]{4.818pt}{0.400pt}}
\put(434.0,119.0){\rule[-0.200pt]{4.818pt}{0.400pt}}
\put(500.0,116.0){\rule[-0.200pt]{0.400pt}{2.168pt}}
\put(490.0,116.0){\rule[-0.200pt]{4.818pt}{0.400pt}}
\put(490.0,125.0){\rule[-0.200pt]{4.818pt}{0.400pt}}
\put(556.0,123.0){\rule[-0.200pt]{0.400pt}{3.613pt}}
\put(546.0,123.0){\rule[-0.200pt]{4.818pt}{0.400pt}}
\put(546.0,138.0){\rule[-0.200pt]{4.818pt}{0.400pt}}
\put(612.0,114.0){\rule[-0.200pt]{0.400pt}{1.445pt}}
\put(602.0,114.0){\rule[-0.200pt]{4.818pt}{0.400pt}}
\put(602.0,120.0){\rule[-0.200pt]{4.818pt}{0.400pt}}
\put(668.0,115.0){\rule[-0.200pt]{0.400pt}{2.650pt}}
\put(658.0,115.0){\rule[-0.200pt]{4.818pt}{0.400pt}}
\put(658.0,126.0){\rule[-0.200pt]{4.818pt}{0.400pt}}
\put(724.0,115.0){\rule[-0.200pt]{0.400pt}{3.613pt}}
\put(714.0,115.0){\rule[-0.200pt]{4.818pt}{0.400pt}}
\put(714.0,130.0){\rule[-0.200pt]{4.818pt}{0.400pt}}
\put(781.0,119.0){\rule[-0.200pt]{0.400pt}{3.132pt}}
\put(771.0,119.0){\rule[-0.200pt]{4.818pt}{0.400pt}}
\put(771.0,132.0){\rule[-0.200pt]{4.818pt}{0.400pt}}
\put(837.0,116.0){\rule[-0.200pt]{0.400pt}{2.409pt}}
\put(827.0,116.0){\rule[-0.200pt]{4.818pt}{0.400pt}}
\put(827.0,126.0){\rule[-0.200pt]{4.818pt}{0.400pt}}
\put(893.0,114.0){\rule[-0.200pt]{0.400pt}{1.445pt}}
\put(883.0,114.0){\rule[-0.200pt]{4.818pt}{0.400pt}}
\put(883.0,120.0){\rule[-0.200pt]{4.818pt}{0.400pt}}
\put(949.0,116.0){\rule[-0.200pt]{0.400pt}{1.686pt}}
\put(939.0,116.0){\rule[-0.200pt]{4.818pt}{0.400pt}}
\put(939.0,123.0){\rule[-0.200pt]{4.818pt}{0.400pt}}
\put(1005.0,121.0){\rule[-0.200pt]{0.400pt}{2.650pt}}
\put(995.0,121.0){\rule[-0.200pt]{4.818pt}{0.400pt}}
\put(995.0,132.0){\rule[-0.200pt]{4.818pt}{0.400pt}}
\put(1061.0,114.0){\rule[-0.200pt]{0.400pt}{1.686pt}}
\put(1051.0,114.0){\rule[-0.200pt]{4.818pt}{0.400pt}}
\put(1051.0,121.0){\rule[-0.200pt]{4.818pt}{0.400pt}}
\put(1117.0,113.0){\rule[-0.200pt]{0.400pt}{0.482pt}}
\put(1107.0,113.0){\rule[-0.200pt]{4.818pt}{0.400pt}}
\put(1107.0,115.0){\rule[-0.200pt]{4.818pt}{0.400pt}}
\put(1173.0,118.0){\rule[-0.200pt]{0.400pt}{2.168pt}}
\put(1163.0,118.0){\rule[-0.200pt]{4.818pt}{0.400pt}}
\put(1163.0,127.0){\rule[-0.200pt]{4.818pt}{0.400pt}}
\put(332.0,754.0){\rule[-0.200pt]{0.400pt}{2.650pt}}
\put(322.0,754.0){\rule[-0.200pt]{4.818pt}{0.400pt}}
\put(322.0,765.0){\rule[-0.200pt]{4.818pt}{0.400pt}}
\put(388.0,747.0){\rule[-0.200pt]{0.400pt}{2.891pt}}
\put(378.0,747.0){\rule[-0.200pt]{4.818pt}{0.400pt}}
\put(378.0,759.0){\rule[-0.200pt]{4.818pt}{0.400pt}}
\put(444.0,735.0){\rule[-0.200pt]{0.400pt}{3.854pt}}
\put(434.0,735.0){\rule[-0.200pt]{4.818pt}{0.400pt}}
\put(434.0,751.0){\rule[-0.200pt]{4.818pt}{0.400pt}}
\put(500.0,715.0){\rule[-0.200pt]{0.400pt}{5.300pt}}
\put(490.0,715.0){\rule[-0.200pt]{4.818pt}{0.400pt}}
\put(490.0,737.0){\rule[-0.200pt]{4.818pt}{0.400pt}}
\put(556.0,702.0){\rule[-0.200pt]{0.400pt}{5.059pt}}
\put(546.0,702.0){\rule[-0.200pt]{4.818pt}{0.400pt}}
\put(546.0,723.0){\rule[-0.200pt]{4.818pt}{0.400pt}}
\put(612.0,668.0){\rule[-0.200pt]{0.400pt}{6.022pt}}
\put(602.0,668.0){\rule[-0.200pt]{4.818pt}{0.400pt}}
\put(602.0,693.0){\rule[-0.200pt]{4.818pt}{0.400pt}}
\put(668.0,640.0){\rule[-0.200pt]{0.400pt}{8.672pt}}
\put(658.0,640.0){\rule[-0.200pt]{4.818pt}{0.400pt}}
\put(658.0,676.0){\rule[-0.200pt]{4.818pt}{0.400pt}}
\put(724.0,623.0){\rule[-0.200pt]{0.400pt}{7.709pt}}
\put(714.0,623.0){\rule[-0.200pt]{4.818pt}{0.400pt}}
\put(714.0,655.0){\rule[-0.200pt]{4.818pt}{0.400pt}}
\put(781.0,602.0){\rule[-0.200pt]{0.400pt}{8.191pt}}
\put(771.0,602.0){\rule[-0.200pt]{4.818pt}{0.400pt}}
\put(771.0,636.0){\rule[-0.200pt]{4.818pt}{0.400pt}}
\put(837.0,585.0){\rule[-0.200pt]{0.400pt}{9.877pt}}
\put(827.0,585.0){\rule[-0.200pt]{4.818pt}{0.400pt}}
\put(827.0,626.0){\rule[-0.200pt]{4.818pt}{0.400pt}}
\put(893.0,542.0){\rule[-0.200pt]{0.400pt}{10.841pt}}
\put(883.0,542.0){\rule[-0.200pt]{4.818pt}{0.400pt}}
\put(883.0,587.0){\rule[-0.200pt]{4.818pt}{0.400pt}}
\put(949.0,514.0){\rule[-0.200pt]{0.400pt}{11.081pt}}
\put(939.0,514.0){\rule[-0.200pt]{4.818pt}{0.400pt}}
\put(939.0,560.0){\rule[-0.200pt]{4.818pt}{0.400pt}}
\put(1005.0,505.0){\rule[-0.200pt]{0.400pt}{10.600pt}}
\put(995.0,505.0){\rule[-0.200pt]{4.818pt}{0.400pt}}
\put(995.0,549.0){\rule[-0.200pt]{4.818pt}{0.400pt}}
\put(1061.0,457.0){\rule[-0.200pt]{0.400pt}{10.118pt}}
\put(1051.0,457.0){\rule[-0.200pt]{4.818pt}{0.400pt}}
\put(1051.0,499.0){\rule[-0.200pt]{4.818pt}{0.400pt}}
\put(1117.0,380.0){\rule[-0.200pt]{0.400pt}{9.395pt}}
\put(1107.0,380.0){\rule[-0.200pt]{4.818pt}{0.400pt}}
\put(1107.0,419.0){\rule[-0.200pt]{4.818pt}{0.400pt}}
\put(1173.0,359.0){\rule[-0.200pt]{0.400pt}{11.322pt}}
\put(1163.0,359.0){\rule[-0.200pt]{4.818pt}{0.400pt}}
\put(1163.0,406.0){\rule[-0.200pt]{4.818pt}{0.400pt}}
\end{picture}
\caption{Probability $P_{\mbox{net}}$ for  the occurence of a
 nontrivial network in the hot and cold phase at the transition point as
function of $\lambda$ for lattice sizes $8^4$ (circles) and$16^4$ (crosses).}
\end{figure}
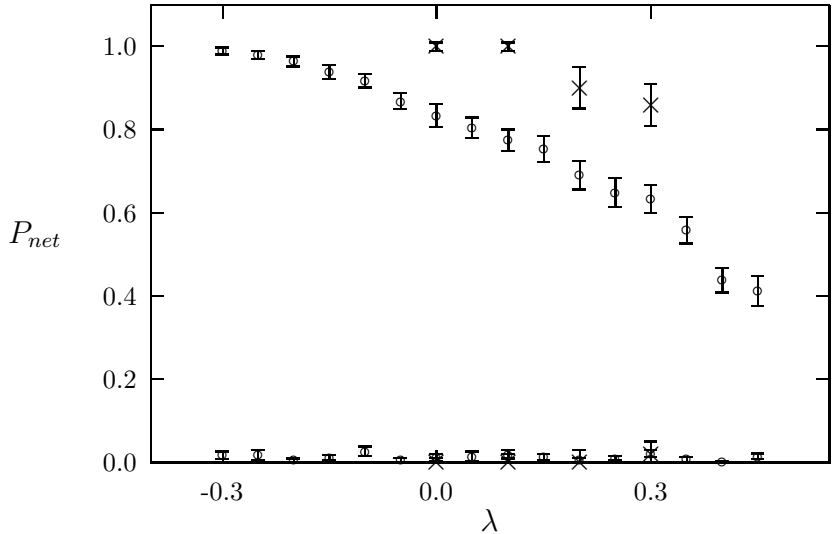

For $\lambda=0$ the events with $\vec{f}\ne 0$ are rare; for lattice
sizes $L=8$ and $L=16$ their percentage is less than 2\% and 0.5\%,
respectively.  For positive $\lambda$ they get slightly more frequent
\cite{krw94}. As is obvious from the pair representation of the
bouquet matrix in Section 3, the net current flow $\vec{f}$ is not
central for the topological characterization. In this context we
should note that the quantity introduced in \cite{bls93} to classify
the networks, because of current conservation, is equal to $\vec{f}$.

For minimizing the number of Dirac plaquettes via a sequence of gauge
transformations we have used an annealing technique \cite{kgv83}.  Our
procedure uses Metropolis sweeps based on a probability distribution 
$P(\eta)\sim\exp(-\alpha\eta)$, where
$\eta=\sum_{x,\mu>\nu}|n_{\mu\nu,x}|$, implemented through random
local gauge transformations with angles maintained within
$[-\pi,\pi)$.  In subsequent simulations we increase the parameter
$\alpha$ in appropriately chosen steps. First we determine $\eta$ as
function of $\alpha$ (which changes little with $L$ and $\lambda$)
using tiny step sizes and large sweep numbers. Then we continue the
procedure with 45 values of $\alpha$ (chosen in the interval
[0.03,100] in such a way as to get approximately equal changes in
$\eta$), performing up to 100 sweeps for each $\alpha$. To be sure
about the results, for each configuration we have generally performed
three such runs as well as some additional runs with many more
sweeps. Repeated applications of the method, with different random
numbers, lead to consistent values for minimal numbers of Dirac
plaquettes, with deviations smaller than 2\%. Thus our procedure
appears to produce a rather reliable determination of the absolute
minimum.

We found that the careful annealing described above is absolutely necessary. 
A minimization of the number of Dirac plaquettes based only on iterative gauge 
transformations gives strongly fluctuating results, which on the average are 
roughly factors 2 and 1.3 larger in the cold and hot phase, respectively,
than the minimal numbers obtained by our procedure. This is not only 
inappropriate in principle but also not sufficient for a proper computational 
identification of the topological structures.

Before applying the annealing procedure the numbers of Dirac
plaquettes are similar in both phases, but the numbers become
markedly different after annealing. For example, for
$L=8$ and $\lambda=0$, in typical configurations with 8703
and 8793 plaquettes in the hot and cold phase, respectively, our
procedure reduced the numbers to 306 and 846.  The stronger reduction
in the cold phase is mainly due to the disappearance of closed
trivial structures. We also note that the procedure not only
affects sheets separately, but involves the full set of them. 
For example, even pairs of oppositely oriented nontrivial sheets,
first observed in \cite{bmmp93}, disappear with our annealing procedure.

Our analysis of minimal Dirac sheet structures leads to results which
agree with those derived from the study of current networks. The hot
phase is characterized by the existence of a Dirac sheet which is
nontrivial in all directions and the cold phase by its absence. To
understand the significance of this result it is important to remember
that, given a nontrivial current network, the sheets could be trivial
or nontrivial (as illustrated by the example of Fig.~1). Thus our
results show that in the hot phase the placement of the largest sheet
is one which makes it topologically nontrivial. This points to the
remarkable fact that the (equivalence classes of) Dirac sheet
structures are the physically relevant objects.

In this context it should be remembered that some time ago Gr\"osch et 
al.~\cite{g-r85} were able to identify nontrivial Dirac sheets without 
boundaries as the structures responsible for the metastable states occurring 
under certain conditions in the cold phase. Here we have shown that 
nontrivial sheet structures with boundaries characterize the hot phase.

To study the role of connectedness, we construct minimal Dirac sheets
by making first the connections where only two Dirac plaquettes meet
at an edge with appropriate orientation. We note then that making
either all or none of the connections where more than two plaquettes
meet (with correct orientation) changes the numbers of sheets and networks, 
and the sizes of the largest sheets, only by 10\% or less and does
not affect the topology. Looking into more detail we find that, for
the most part, these changes are due to intersections of the type of
Figs.~2a and 2b, where the sheets must actually be kept separated.
Taking this into account the change reduces to 3\% or less. We thus
obtain the further remarkable result that intersections of minimal
sheets are not important.

\begin{figure}
\setlength{\unitlength}{0.240900pt}
\ifx\plotpoint\undefined\newsavebox{\plotpoint}\fi
\begin{picture}(1349,900)(0,0)
\font\gnuplot=cmr10 at 10pt
\gnuplot
\sbox{\plotpoint}{\rule[-0.200pt]{0.400pt}{0.400pt}}%
\put(220.0,676.0){\rule[-0.200pt]{4.818pt}{0.400pt}}
\put(198,676){\makebox(0,0)[r]{3600}}
\put(1265.0,676.0){\rule[-0.200pt]{4.818pt}{0.400pt}}
\put(220.0,488.0){\rule[-0.200pt]{4.818pt}{0.400pt}}
\put(198,488){\makebox(0,0)[r]{2400}}
\put(1265.0,488.0){\rule[-0.200pt]{4.818pt}{0.400pt}}
\put(220.0,301.0){\rule[-0.200pt]{4.818pt}{0.400pt}}
\put(198,301){\makebox(0,0)[r]{1200}}
\put(1265.0,301.0){\rule[-0.200pt]{4.818pt}{0.400pt}}
\put(220.0,113.0){\rule[-0.200pt]{4.818pt}{0.400pt}}
\put(198,113){\makebox(0,0)[r]{0}}
\put(1265.0,113.0){\rule[-0.200pt]{4.818pt}{0.400pt}}
\put(332.0,113.0){\rule[-0.200pt]{0.400pt}{4.818pt}}
\put(332,68){\makebox(0,0){-0.3}}
\put(332.0,812.0){\rule[-0.200pt]{0.400pt}{4.818pt}}
\put(668.0,113.0){\rule[-0.200pt]{0.400pt}{4.818pt}}
\put(668,68){\makebox(0,0){0.0}}
\put(668.0,812.0){\rule[-0.200pt]{0.400pt}{4.818pt}}
\put(1005.0,113.0){\rule[-0.200pt]{0.400pt}{4.818pt}}
\put(1005,68){\makebox(0,0){0.3}}
\put(1005.0,812.0){\rule[-0.200pt]{0.400pt}{4.818pt}}
\put(220.0,113.0){\rule[-0.200pt]{256.559pt}{0.400pt}}
\put(1285.0,113.0){\rule[-0.200pt]{0.400pt}{173.207pt}}
\put(220.0,832.0){\rule[-0.200pt]{256.559pt}{0.400pt}}
\put(45,472){\makebox(0,0){ }}
\put(752,23){\makebox(0,0){$\lambda$ }}
\put(752,877){\makebox(0,0){ }}
\put(1307,301){\makebox(0,0)[l]{200}}
\put(1307,488){\makebox(0,0)[l]{400}}
\put(1307,676){\makebox(0,0)[l]{600}}
\put(758,676){\makebox(0,0)[l]{hot(sheet)}}
\put(668,347){\makebox(0,0)[l]{hot(network)}}
\put(332,183){\makebox(0,0)[l]{cold}}
\put(186,754){\makebox(0,0)[r]{links}}
\put(1307,754){\makebox(0,0)[l]{plaq.}}
\put(1307,113){\makebox(0,0)[l]{0}}
\put(220.0,113.0){\rule[-0.200pt]{0.400pt}{173.207pt}}
\put(332,143){\circle{12}}
\put(388,144){\circle{12}}
\put(444,144){\circle{12}}
\put(500,144){\circle{12}}
\put(556,151){\circle{12}}
\put(612,146){\circle{12}}
\put(668,147){\circle{12}}
\put(724,144){\circle{12}}
\put(781,148){\circle{12}}
\put(837,147){\circle{12}}
\put(893,143){\circle{12}}
\put(949,142){\circle{12}}
\put(1005,145){\circle{12}}
\put(1061,142){\circle{12}}
\put(1117,141){\circle{12}}
\put(1173,142){\circle{12}}
\put(332,369){\circle{12}}
\put(388,354){\circle{12}}
\put(444,336){\circle{12}}
\put(500,324){\circle{12}}
\put(556,306){\circle{12}}
\put(612,294){\circle{12}}
\put(668,281){\circle{12}}
\put(724,269){\circle{12}}
\put(781,262){\circle{12}}
\put(837,253){\circle{12}}
\put(893,248){\circle{12}}
\put(949,237){\circle{12}}
\put(1005,235){\circle{12}}
\put(1061,226){\circle{12}}
\put(1117,214){\circle{12}}
\put(1173,208){\circle{12}}
\put(332.0,141.0){\rule[-0.200pt]{0.400pt}{0.964pt}}
\put(322.0,141.0){\rule[-0.200pt]{4.818pt}{0.400pt}}
\put(322.0,145.0){\rule[-0.200pt]{4.818pt}{0.400pt}}
\put(388.0,141.0){\rule[-0.200pt]{0.400pt}{1.445pt}}
\put(378.0,141.0){\rule[-0.200pt]{4.818pt}{0.400pt}}
\put(378.0,147.0){\rule[-0.200pt]{4.818pt}{0.400pt}}
\put(444.0,142.0){\rule[-0.200pt]{0.400pt}{0.964pt}}
\put(434.0,142.0){\rule[-0.200pt]{4.818pt}{0.400pt}}
\put(434.0,146.0){\rule[-0.200pt]{4.818pt}{0.400pt}}
\put(500.0,142.0){\rule[-0.200pt]{0.400pt}{0.964pt}}
\put(490.0,142.0){\rule[-0.200pt]{4.818pt}{0.400pt}}
\put(490.0,146.0){\rule[-0.200pt]{4.818pt}{0.400pt}}
\put(556.0,149.0){\rule[-0.200pt]{0.400pt}{0.964pt}}
\put(546.0,149.0){\rule[-0.200pt]{4.818pt}{0.400pt}}
\put(546.0,153.0){\rule[-0.200pt]{4.818pt}{0.400pt}}
\put(612.0,144.0){\rule[-0.200pt]{0.400pt}{0.964pt}}
\put(602.0,144.0){\rule[-0.200pt]{4.818pt}{0.400pt}}
\put(602.0,148.0){\rule[-0.200pt]{4.818pt}{0.400pt}}
\put(668.0,144.0){\rule[-0.200pt]{0.400pt}{1.204pt}}
\put(658.0,144.0){\rule[-0.200pt]{4.818pt}{0.400pt}}
\put(658.0,149.0){\rule[-0.200pt]{4.818pt}{0.400pt}}
\put(724.0,143.0){\rule[-0.200pt]{0.400pt}{0.723pt}}
\put(714.0,143.0){\rule[-0.200pt]{4.818pt}{0.400pt}}
\put(714.0,146.0){\rule[-0.200pt]{4.818pt}{0.400pt}}
\put(781.0,147.0){\rule[-0.200pt]{0.400pt}{0.723pt}}
\put(771.0,147.0){\rule[-0.200pt]{4.818pt}{0.400pt}}
\put(771.0,150.0){\rule[-0.200pt]{4.818pt}{0.400pt}}
\put(837.0,146.0){\rule[-0.200pt]{0.400pt}{0.723pt}}
\put(827.0,146.0){\rule[-0.200pt]{4.818pt}{0.400pt}}
\put(827.0,149.0){\rule[-0.200pt]{4.818pt}{0.400pt}}
\put(893.0,142.0){\rule[-0.200pt]{0.400pt}{0.723pt}}
\put(883.0,142.0){\rule[-0.200pt]{4.818pt}{0.400pt}}
\put(883.0,145.0){\rule[-0.200pt]{4.818pt}{0.400pt}}
\put(949.0,140.0){\rule[-0.200pt]{0.400pt}{0.723pt}}
\put(939.0,140.0){\rule[-0.200pt]{4.818pt}{0.400pt}}
\put(939.0,143.0){\rule[-0.200pt]{4.818pt}{0.400pt}}
\put(1005.0,143.0){\rule[-0.200pt]{0.400pt}{0.964pt}}
\put(995.0,143.0){\rule[-0.200pt]{4.818pt}{0.400pt}}
\put(995.0,147.0){\rule[-0.200pt]{4.818pt}{0.400pt}}
\put(1061.0,141.0){\rule[-0.200pt]{0.400pt}{0.482pt}}
\put(1051.0,141.0){\rule[-0.200pt]{4.818pt}{0.400pt}}
\put(1051.0,143.0){\rule[-0.200pt]{4.818pt}{0.400pt}}
\put(1117.0,139.0){\rule[-0.200pt]{0.400pt}{0.723pt}}
\put(1107.0,139.0){\rule[-0.200pt]{4.818pt}{0.400pt}}
\put(1107.0,142.0){\rule[-0.200pt]{4.818pt}{0.400pt}}
\put(1173.0,140.0){\rule[-0.200pt]{0.400pt}{0.723pt}}
\put(1163.0,140.0){\rule[-0.200pt]{4.818pt}{0.400pt}}
\put(1163.0,143.0){\rule[-0.200pt]{4.818pt}{0.400pt}}
\put(332.0,366.0){\rule[-0.200pt]{0.400pt}{1.445pt}}
\put(322.0,366.0){\rule[-0.200pt]{4.818pt}{0.400pt}}
\put(322.0,372.0){\rule[-0.200pt]{4.818pt}{0.400pt}}
\put(388.0,352.0){\rule[-0.200pt]{0.400pt}{0.964pt}}
\put(378.0,352.0){\rule[-0.200pt]{4.818pt}{0.400pt}}
\put(378.0,356.0){\rule[-0.200pt]{4.818pt}{0.400pt}}
\put(444.0,334.0){\rule[-0.200pt]{0.400pt}{1.204pt}}
\put(434.0,334.0){\rule[-0.200pt]{4.818pt}{0.400pt}}
\put(434.0,339.0){\rule[-0.200pt]{4.818pt}{0.400pt}}
\put(500.0,321.0){\rule[-0.200pt]{0.400pt}{1.204pt}}
\put(490.0,321.0){\rule[-0.200pt]{4.818pt}{0.400pt}}
\put(490.0,326.0){\rule[-0.200pt]{4.818pt}{0.400pt}}
\put(556.0,304.0){\rule[-0.200pt]{0.400pt}{1.204pt}}
\put(546.0,304.0){\rule[-0.200pt]{4.818pt}{0.400pt}}
\put(546.0,309.0){\rule[-0.200pt]{4.818pt}{0.400pt}}
\put(612.0,292.0){\rule[-0.200pt]{0.400pt}{1.204pt}}
\put(602.0,292.0){\rule[-0.200pt]{4.818pt}{0.400pt}}
\put(602.0,297.0){\rule[-0.200pt]{4.818pt}{0.400pt}}
\put(668.0,279.0){\rule[-0.200pt]{0.400pt}{0.964pt}}
\put(658.0,279.0){\rule[-0.200pt]{4.818pt}{0.400pt}}
\put(658.0,283.0){\rule[-0.200pt]{4.818pt}{0.400pt}}
\put(724.0,267.0){\rule[-0.200pt]{0.400pt}{0.964pt}}
\put(714.0,267.0){\rule[-0.200pt]{4.818pt}{0.400pt}}
\put(714.0,271.0){\rule[-0.200pt]{4.818pt}{0.400pt}}
\put(781.0,260.0){\rule[-0.200pt]{0.400pt}{0.964pt}}
\put(771.0,260.0){\rule[-0.200pt]{4.818pt}{0.400pt}}
\put(771.0,264.0){\rule[-0.200pt]{4.818pt}{0.400pt}}
\put(837.0,251.0){\rule[-0.200pt]{0.400pt}{1.204pt}}
\put(827.0,251.0){\rule[-0.200pt]{4.818pt}{0.400pt}}
\put(827.0,256.0){\rule[-0.200pt]{4.818pt}{0.400pt}}
\put(893.0,246.0){\rule[-0.200pt]{0.400pt}{0.964pt}}
\put(883.0,246.0){\rule[-0.200pt]{4.818pt}{0.400pt}}
\put(883.0,250.0){\rule[-0.200pt]{4.818pt}{0.400pt}}
\put(949.0,235.0){\rule[-0.200pt]{0.400pt}{0.964pt}}
\put(939.0,235.0){\rule[-0.200pt]{4.818pt}{0.400pt}}
\put(939.0,239.0){\rule[-0.200pt]{4.818pt}{0.400pt}}
\put(1005.0,233.0){\rule[-0.200pt]{0.400pt}{1.204pt}}
\put(995.0,233.0){\rule[-0.200pt]{4.818pt}{0.400pt}}
\put(995.0,238.0){\rule[-0.200pt]{4.818pt}{0.400pt}}
\put(1061.0,224.0){\rule[-0.200pt]{0.400pt}{1.204pt}}
\put(1051.0,224.0){\rule[-0.200pt]{4.818pt}{0.400pt}}
\put(1051.0,229.0){\rule[-0.200pt]{4.818pt}{0.400pt}}
\put(1117.0,212.0){\rule[-0.200pt]{0.400pt}{0.964pt}}
\put(1107.0,212.0){\rule[-0.200pt]{4.818pt}{0.400pt}}
\put(1107.0,216.0){\rule[-0.200pt]{4.818pt}{0.400pt}}
\put(1173.0,205.0){\rule[-0.200pt]{0.400pt}{1.204pt}}
\put(1163.0,205.0){\rule[-0.200pt]{4.818pt}{0.400pt}}
\put(1163.0,210.0){\rule[-0.200pt]{4.818pt}{0.400pt}}
\put(332,142){\makebox(0,0){$\times$}}
\put(668,149){\makebox(0,0){$\times$}}
\put(1005,187){\makebox(0,0){$\times$}}
\put(332,729){\makebox(0,0){$\times$}}
\put(668,678){\makebox(0,0){$\times$}}
\put(1005,535){\makebox(0,0){$\times$}}
\put(332.0,139.0){\rule[-0.200pt]{0.400pt}{1.204pt}}
\put(322.0,139.0){\rule[-0.200pt]{4.818pt}{0.400pt}}
\put(322.0,144.0){\rule[-0.200pt]{4.818pt}{0.400pt}}
\put(668.0,143.0){\rule[-0.200pt]{0.400pt}{2.650pt}}
\put(658.0,143.0){\rule[-0.200pt]{4.818pt}{0.400pt}}
\put(658.0,154.0){\rule[-0.200pt]{4.818pt}{0.400pt}}
\put(1005.0,171.0){\rule[-0.200pt]{0.400pt}{7.709pt}}
\put(995.0,171.0){\rule[-0.200pt]{4.818pt}{0.400pt}}
\put(995.0,203.0){\rule[-0.200pt]{4.818pt}{0.400pt}}
\put(332.0,702.0){\rule[-0.200pt]{0.400pt}{12.768pt}}
\put(322.0,702.0){\rule[-0.200pt]{4.818pt}{0.400pt}}
\put(322.0,755.0){\rule[-0.200pt]{4.818pt}{0.400pt}}
\put(668.0,649.0){\rule[-0.200pt]{0.400pt}{13.731pt}}
\put(658.0,649.0){\rule[-0.200pt]{4.818pt}{0.400pt}}
\put(658.0,706.0){\rule[-0.200pt]{4.818pt}{0.400pt}}
\put(1005.0,507.0){\rule[-0.200pt]{0.400pt}{13.490pt}}
\put(995.0,507.0){\rule[-0.200pt]{4.818pt}{0.400pt}}
\put(995.0,563.0){\rule[-0.200pt]{4.818pt}{0.400pt}}
\end{picture}
\caption{Size of largest current network (circles, links) and largest
 minimal Dirac sheet (crosses, plaquettes) in the hot and cold phase 
 at the transition point, as function of $\lambda$ on $8^4$ lattice.}
\end{figure}
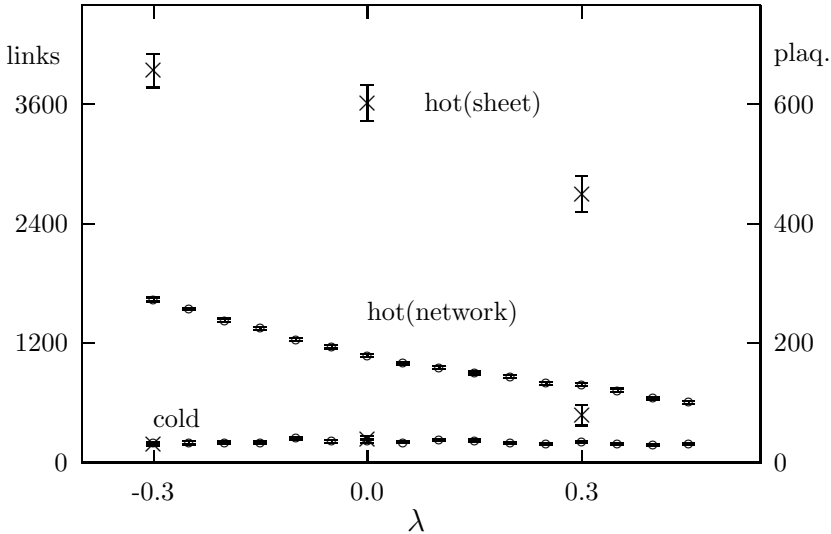

In Fig.~4 we present the sizes of the largest current network and of
the largest minimal Dirac sheet. Obviously the sheets give a more
sensitive signal for the different phases. This may be explained by
the fact that area instead of length enters. It is to be noted
that the definition we have used for the connectedness of sheets
plays an important role in this result.  If one considers as 
connected sheets which have just sites in common, then the signal
provided by the size of the largest minimal sheets becomes 
comparable to the one given by the networks.

Fig.~5 indicates that the probability $P_{\mbox{\scriptsize net}}$ of
the occurrence of a non-trivial network, taking values very close
1 or 0 for hot and cold phase respectively, may be a more sensitive
order parameter than $n_{\mbox{\scriptsize max}}/n_{\mbox{\scriptsize
tot}}$, the relative size of the largest network, advocated in
\cite{bfk94}.

\begin{figure}[tb]
\begin{center}
\leavevmode
\psfig{figure=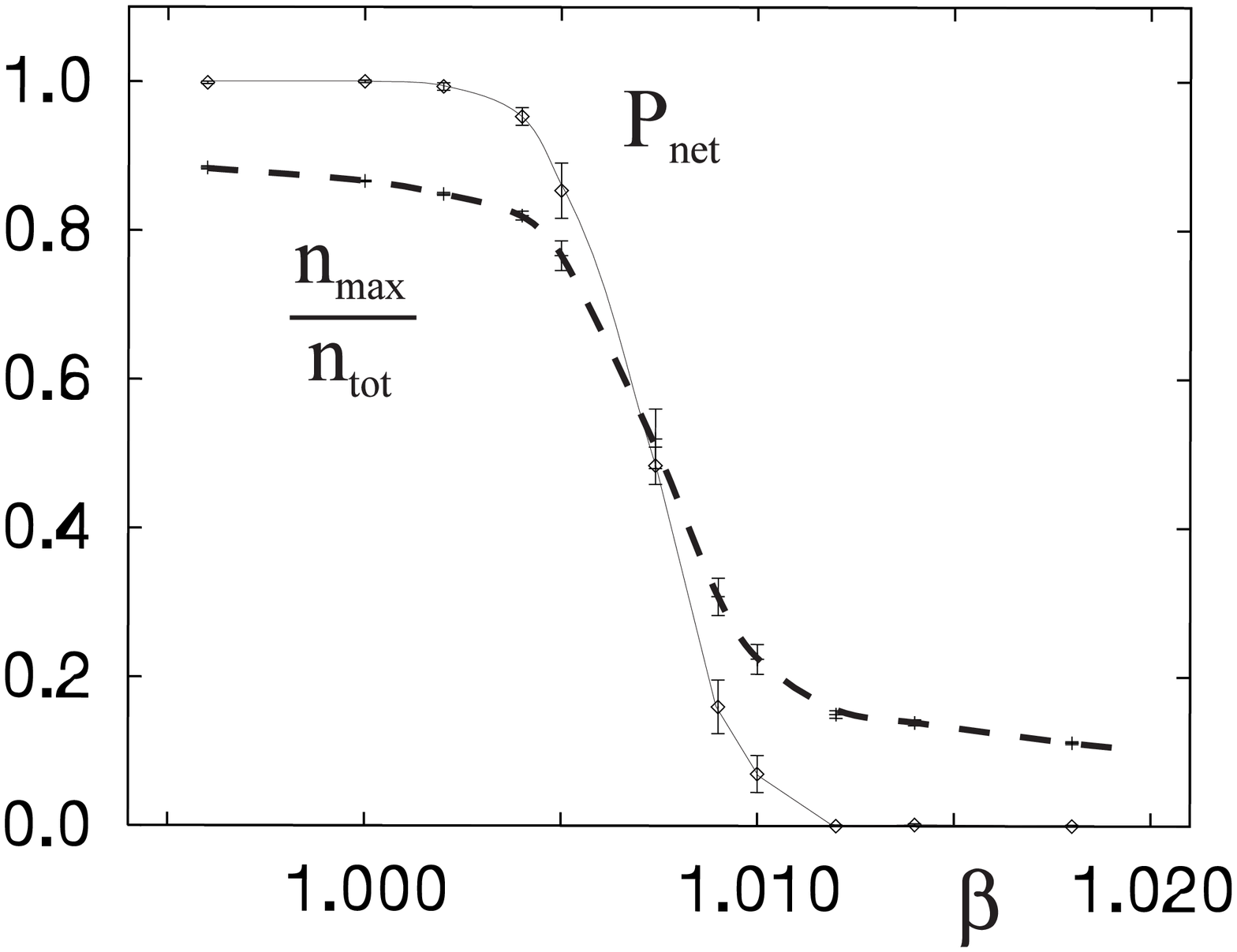,bbllx=2095bp,bblly=2132bp,bburx=2674bp,%
       bbury=2575bp,width=10cm,angle=0}
\end{center}
\caption{Order parameters $P_{\mbox{net}}$ and 
$n_{\mbox{max}}/ n_{\mbox{tot}}$ for $\lambda=0$ 
as functions
of $\beta$ on $8^4$ lattice.}
\end{figure}
\clearpage

\section{Discussion}
\hspace{3mm}
The success of our topological order parameter confirms that the 
phase transition of the 4-dimensional compact U(1) gauge field theory 
concides with a percolation-type transition of the topological structures.
Examples of theories with different behaviors have recently be discussed in 
\cite{bfk94}.

In order to see the general meaning of this results we note that a 
topologically nontrivial structure on a finite lattice with periodic boundary 
conditions, i.e. on the torus $\mbox{\bf T}^4$, corresponds to an infinite 
structure on an infinite lattice. Thus the hot phase is more generally 
characterized by the existence of an infinite current network or Dirac sheet 
and the cold phase by its absence, where on finite lattices 
``infinite'' is to be defined in accordance with the particular boundary 
conditions which are used. 

To test this characterization we also performed simulations with open boundary 
conditions \cite{krw95}. In the corresponding analysis the prescription 
``nontrivial in all directions'' is replaced by ``touching the boundary in 
all directions''. In this way we again get an unambiguous signal for the
phases. The comparison of the order parameters looks very similar to Fig.~5, 
the advantage of our order parameter being even more pronounced. 
Because of finite-size effects, however, the width of transition region  
becomes larger, for $L=16$, by a factor of approximately $100$ . 

In order to obtain information about the order of the phase transition 
the most immediate thing to do is to look whether there is an energy gap 
for sufficiently large $L$.  With periodic boundaries $L=8$ is sufficient 
for seeing a gap whereas $L=4$ is not. With open boundaries a gap is not yet 
seen for $L=16$. The width of the peak is relatively narrow (comparable 
to the width of the peaks with $L=8$ in the periodic-boundary case) 
\cite{krw95}. It appears to us that lattices well larger than $L=16$
will be needed in order to see a possible gap in systems with open boundary
conditions.

In \cite{ln93} the simulations have been performed on the surface of a 5 
dimensional cube which is homeomorphic to the sphere $\mbox{\bf S}^4$. 
Considering lattices up to $L=10$ the authors observe only one peak. Its width
for $L=10$ is comparable to those of the peaks on $\mbox{\bf T}^4$ for $L=8$. 
The extension of the transition region, which could also provide information
about finite-size effects, has not been reported.

In \cite{bf94} simulations for lattices up to $L=16$ have been done
with fixed boundaries (i.e.~setting the group elements to {\bf 1} at
the boundaries), which makes the lattice again homeomorphic to
$\mbox{\bf S}^4$. With these boundary conditions the transition region
was observed to be wider by a factor of about $10^3$, which reflects
the huge finite-size effects caused by the strong inhomogeneity of the
system.  Because of low statistics the results in this work are of
semiqualitative nature and the energy distribution is not given.

It seems to us that in the two geometries homeomorphic to 
$\mbox{\bf S}^4$ mentioned above larger lattices would also be  necessary 
for deciding about the existence of a gap. Further, in these geometries 
the analysis of configurations of monopole currents and Dirac sheets 
would be of great interest, too.  Appropriate definitions of 
``infinite'' on finite lattices should of course be given. It should 
be realized that these geometries are different from a physical point of view.
Fixed boundary conditions can be expected to lead to the same (flat-space) 
limit as periodic or open boundary conditions, while for the surface of a 
5-dimensional hypercube some curvature (depending on the details of the limit)
could persist. 

A further type of boundary condition, proposed in \cite{bls94},
consists in the suppression of (all) monopoles at the boundaries. This
is done by using an action leading to the suppression of monopoles
\cite{bmm93} at the boundaries and the usual Wilson action
elsewhere. Though the definition of ``infinite'' gets somewhat dilute
in that case we expect it to fit into our picture as well. The
existence of a gap with this type of boundary condition remains to be
confirmed.

\section*{Acknowledgements}
\hspace{3mm}
This research was supported in part under DFG grants Ke 250/7-2 and 250/11-1 
and under DOE grant DE-FG02-91ER40676.
The computations were done on the CM5 of the Center for Computational
Science of Boston University and on the CM5 of the GMD at St.~Augustin.

\end{document}